\documentstyle[12pt]{article}
\begin {document}
\begin {center}
{\Large\bf ON THE QUESTION OF DEGENERACY OF TOPOLOGICAL SOLITONS IN A
GAUGED O(3) NON-LINEAR SIGMA MODEL WITH CHERN-SIMONS TERM}
\vskip 1in
{\bf P.Mukherjee}\footnote{e-mail:pradip@boson.bose.res.in}\\
{\normalsize Department of Physics\\
A.B.N.Seal College,Coochbehar\\
West Bengal,India}
\end {center}
\begin {abstract}
We show that the  degeneracy of topological solitons
in the gauged O(3) non-linear
sigma model with Chern-Simons term  may be removed by chosing a
self-interaction potential with symmetry - breaking minima.
The topological solitons in the model has energy,charge,flux
and angular momentum quantised in each topological sector.
\end {abstract}
\newpage
The 2+1 dimensional O(3) nonlinear sigma model has been studied
extensively over a long period of time[1] due to its own interest of
providing topologically stable soliton solutions which are exactly
integrable in the Bogomol'nyi limit [2] and also for its applications
in condensed matter physics [3].Soliton solutions of the model modified
by the addition of Hopf term characterising maps from $S^3$ to $S^2$
reveal the occurence of fractional spin and statistics [4,5]. The system
can be cast in the form of a genuine gauge theory by the inclusion of the
Chern - Simons (C-S) term which implements
fractional spin and statistics in the
context of local field theories .A gauge-independent analysis [6] of the
O(3) sigma model with C-S term [7] shows conclusively that this
fractional spin is a physical effect and not an artifact of the gauge.The
fractional spin of the topologically stable solitons of the model were
shown to scale as the square of the vortex number .It is also established
that this property is not specific of the particular model [8-11] and rather
shared by the Chern-Simons vortices in general [12].

A characteristic feature of the soliton solutions of the
O(3) sigma model in (2+1) dimensions is its scale
- invariance which prevents particle interpretation on quantisation [13].
An interesting method of breaking this scale-invariance is to gauge the U(1)
subgroup as well as including a potential term [14].Recently it has been
shown  that gauging the U(1)
subgroup by Chern - Simons term with a particlar
potential gives rise to both topological and nontopological  solitons[15].
However the topological solitons are infinitely degenerate in a given topo-
logical sector with quantised energy but degenerate charge , flux and
angular momentum .This is certainly very characteristic outcome
in the light of the findings detailed above about the Chern - Simons vortices.

The potential used in [15] has two discrete minima $\phi_3 = \pm 1$
and the U(1) symmetry is not spontaneously broken.The topologically stable
soliton solutions of the model are classified according to the homotopy
$\Pi_2(S_2)=Z$ just as the model without the gauge field coupling.The observed
infinite degeneracy in each topological sector is thus physically undesirable
.In the present letter we will show that inclusion of a different form of
self-interaction with symmetry breaking minima leads to topologically stable
soliton solutions which have all the desired features with quantised
energy,charge,flux and angular momentum in each topological sector.

The Lagrangian of our model is given by
\begin {equation}
{\cal L} ={1 \over 2}D_\mu {\bf \phi}\cdot D^\mu{\bf \phi}
	 +{k \over 4}\epsilon^{\mu\nu\lambda}A_\mu\partial_\nu A_\lambda+
	   U({\bf \phi})
\end {equation}
Here ${\bf \phi}$ is a triplet of scalar fields constituting a vector in the
internal space with unit norm
\begin {eqnarray}
\phi_a = {\bf n_a}\cdot {\bf \phi},(a=1,2,3)\\
{\bf \phi}\cdot{\bf \phi} =\phi_a\phi_a= 1
\end {eqnarray}
where ${\bf  n}_a$
constitute a basis of unit orthogonal vectors in the internal space.
We work in the Minkowskian space - time with the metric tensor
diagonal, $g_{\mu\nu} = (1,-1,-1)$.

$D_\mu {\bf \phi}$ is the covariant derivative given by
\begin {equation}
D_\mu {\bf{\phi}}=\partial_\mu {\bf{\phi}}
   + A_\mu {\bf n}_3\times{\bf{\phi}}
\end {equation}
 The SO(2)
(U(1)) subgroup is gauged by the vector potential $A_\mu$ whose dynamics is
 dictated
by the CS term.The potential
\begin {equation}
U({\bf{\phi}})=-{1 \over {2k^2}} \phi_3^2(1-\phi_3^2)
\end {equation}
gives a self interaction of the fields $\phi_a$.Note that the minima
of the potential arise when either,
\begin {eqnarray}
\phi_1 =& \phi_2 = 0\hspace{.2cm} and \hspace{.2cm}\phi_3 = \pm 1\\
or,\phi_3 =& 0\hspace{.2cm} and \hspace{.2cm}\phi_1^2+\phi_2^2 =1
\end {eqnarray}
In (6) the U(1) symmetry is unbroken whereas (7) corresponds to the spontaneous
breaking of the same.For obvious reasons we will refer to (6) as the symmetric
minima and (7) as the symmetry breaking minima.

The Euler - Lagrange equations of the system (1) is derived subject to the
constraint (3) by the Lagrange multiplier technique
\begin {eqnarray}
D_\nu (D^\nu {\bf{ \phi}})& =& [D_\nu (D^\nu {\bf{ \phi}})
\cdot{\bf{ \phi}}]{\bf{ \phi}}
   -{1\over k^2}{\bf n}_3\phi_3(1-2\phi_3^2)
	      +{1\over k^2}\phi_3^2(1-2\phi_3^2){\bf \phi}\\
{k \over 2}\epsilon^{\mu\nu\lambda}F_{\nu\lambda}& =& j^\mu
\end {eqnarray}
where
\begin {equation}
j^\mu = -{\bf n}_3\cdot{\bf J}^\mu\hspace{.2cm} and\hspace{.2cm}
 {\bf J}^\mu ={\bf{ \phi}}\times D^\mu {\bf{ \phi}}
 \end {equation}
Using (8) we get
\begin {equation}
D_\mu {\bf J}^\mu = {1 \over k^2}({\bf n}_3
\times {\bf {\phi}})\phi_3(1-2\phi_3^2)
\end {equation}

From (9) we find
\begin {equation}
j^0 = k\epsilon_{ij}\partial^iA^j = - kB
\end {equation}
where B = curl{\bf A} is the magnetic field. Integrating (12) over
 the entire space
we obtain
\begin {equation}
\Phi = -{Q \over k}
\end {equation}
where Q is the charge and $\Phi$ is the magnetic flux.The relation (13) is
characteristic of the CS theories.

The energy functional is now obtained from Schwinger's energy - momentum tensor
[16] which in the static limit becomes
\begin {equation}
E = {1 \over 2}\int d^2x[(D_i{\bf {\phi}})\cdot(D_i{\bf {\phi}})+
{{k^2B^2}\over{1-\phi_3^2}}+{1 \over k^2}\phi_3^2(1-\phi_3^2)]
\end {equation}
We have eliminated $A_0$ using (10) and (12).The energy functional  (14)
 is subject to the
constraint (3).
We can construct a conserved current
\begin {equation}
K_\mu = {1 \over {8\pi}}\epsilon_{\mu\nu\lambda}[
{\bf {\phi}}\cdot D^\nu {\bf {\phi}}
\times D^\lambda{\bf {\phi}} - F^{\nu\lambda}\phi_3]
\end {equation}
By a straightforward calculation it can be shown that
\begin {equation}
\partial_\mu K^\mu = 0
\end {equation}
The corresponding conserved charge is
\begin {equation}
T = \int d^2x K_0
\end {equation}

Using (15) and (17) we can write

\begin {equation}
T = \int d^2x[{1 \over{8\pi}}\epsilon_{ij}{\bf {\phi}}
\cdot(\partial^i{\bf {\phi}}
\times \partial^j {\bf {\phi}})]
+{ 1 \over {4\pi}}\int_{boundary}\phi_3 A_\theta r d\theta
\end {equation}
where r,$\theta$ are polar coordinates in the physical space and $A_\theta
= {\bf e}_\theta \cdot {\bf A}$.

Let us now consider the symmetric minima (6).For finite value of the energy
 functional (14) we require the fields at the spatial infinity to be equal
to either $\phi_3=1 or -1$.The physical infinity is thus one point compactified
to either the north or the south pole of the internal sphere.The static field
configurations are thus classified according to the degree of the mapping
from $S_2$ to $S_2$.Note that the first term of (18) gives the winding number
of the mapping [3].But $\phi_3 \to \pm 1$ on the boundary.As a result the
value of the topological charge is not quantised.So the static finite energy
solutions corresponding to the symmetric minima are nontopological.Unlike [15]
topological solitons are not obtained in this limit.These observations may
be compared with earlier findings about the Chern - Simons solitons [17].

The situation changes dramatically when we consider the symmetry breaking
minima (7).Here the physical vaccua bear representation of the U(1) symmetry.
\begin {equation}
\psi \approx e^{in\theta}
\end {equation}
where $\psi = \phi_1 + i\phi_2$ and n gives the number of times the infinite
circle of the physical space circuitting around the equatorial circle of the
internal sphere.The topological solitons of the model are now classified
according to this winding number.When the equatorial circle is traversed once
the physical space is mapped on a hemisphere of the internal sphere.In
general the topological charge (18) will be quantised by
\begin {equation}
T ={ n \over 2}
\end {equation}
allowing half integral values of T.

Using the definition of $\phi$ we can write
\begin {equation}
D_i {\bf {\phi}}\cdot D_i{\bf{\phi}}
      = |(\partial_i + iA_i)\psi|^2 +(\partial_i\phi_3)^2
\end {equation}
From (14) and (21) we observe that for finite energy configurations we
require the covariant derivative $(\partial_i + iA_i)\psi$ to vanish at
the physical boundary.Using (19) we then get on the boundary
\begin {equation}
{\bf{A}}={\bf{e_\theta}}{n\over r}
\end {equation}

It is really interesting to observe that the asymptotic form (22) is
sufficient to find the magnetic flux $\Phi$,charge Q and spin S.
Thus the magnetic flux
\begin {equation}
\Phi = \int B d^2x = \int_{boundary}A_\theta r d\theta = 2\pi n
\end {equation}
and spin [6,12]
\begin {equation}
S = -{k\over 2} \int_{boundary}
  \partial^i[x_iA^2 - A_ix_jA^j]d^2x = -\pi kn^2
\end {equation}
Using (13) and (23) we then find
\begin {equation}
Q = -2\pi kn
\end {equation}
Equations (23) to (25) show that the soliton solutions corresponding to the
symmetry breaking vaccua have charge,flux and angular momentum quantised in
each topological sector.

We then turn to show that the model satisfies Bogomol'nyi conditions.
Rearrenging (14) we can write [18]

\begin {equation}
E = {1 \over 2} \int d^{2}x[{1 \over 2}(D_i{\bf {\phi}} \pm
\epsilon_{ij}{\bf {\phi}}\times	 D_{j}{\bf {\phi}})^2 +
{{k^2} \over {1-\phi_3^2}}(F_{12}\pm
{1 \over k^2}\phi_3(1- \phi_3^2))^2]
    \pm 4\pi T
  \end {equation}
  Equation (26) gives the Bogomol'nyi conditions
  \begin {eqnarray}
  D_i{\bf {\phi}}\pm  \epsilon_{ij} {\bf {\phi}}\times D_j{\bf \phi} = 0\\
  F_{12}\pm {1 \over k^2}\phi_3(1-\phi_3^2) =0
 \end {eqnarray}
  which minimises the energy functional in a particular topological sector,
  the upper sign corresponds to +ve and the lower sign corresponds to -ve
  value of the topological charge.The equations can be handled in the usual
method[3,15] to show that the scale invariance is removed by the artifice
of the gauge-field coupling.

  We will now show the consistency of (27) and (28) using the well-known
  Ansatz[14,19]
  \begin {eqnarray}
  \phi_1(r,\theta) = \sin F(r) \cos n\theta\nonumber\\
  \phi_2(r,\theta) = \sin F(r) \sin n\theta\nonumber\\
  \phi_3(r,\theta) = \cos F(r)\nonumber\\
  {\bf A}(r,\theta)= -{\bf e}_\theta {{na(r)} \over r}
  \end {eqnarray}
  From (7) we observe that we require the boundary condition
  \begin {equation}
  F(r) \to \pm {\pi \over 2}\hspace{.2cm}
as\hspace{.2cm} r \to \infty
  \end {equation}
and equation (22) dictates that
  \begin {equation}
a(r) \to -1 \hspace{.2cm}as\hspace{.2cm} r \to \infty
  \end {equation}
Remember that equation (22) was obtained so as the solutions have finite
energy.
  Again ,for the fields to be well defined at the origin we require
  \begin {equation}
  F(r) \to 0 or \pi \hspace{.2cm}and \hspace{.2cm}
  a(r) \to 0\hspace{.2cm} as \hspace{.2cm}r \to 0
  \end {equation}
Substituting the Ansatz(29) into (27) and (28) we find that
  \begin {eqnarray}
  F^\prime (r) = \pm {{n(a+1)}\over r} \sin F\\
  a^\prime (r) = \mp {r \over {nk^2}}\sin^2F \cos F
  \end {eqnarray}
  where the upper sign holds for +ve T and the lower sign corresponds to
  -ve T.Equations (33) and (34) are not exactly integrable.They may be
solved numerically subject to the appropriate boundary conditions to get
the exact profiles.

Using the Ansatz (29) we can explicitly compute the topological charge T
by performing the integration in (18).The result is
\begin {equation}
T = -{n\over 2}[cosf(\infty)-cosf(0)]
\end {equation}
So we find that according to (30) and (32) T =$\pm {n\over 2}$ which is in
agreement with our observation (20).Note that f(0) corresponds to +ve T
and f(0) = $\pi$ corresponds to -ve T.
If we take +ve T we find F(r) bounded between 0 and
  $\pi\over 2$
  is consistent with (30),(32) and (33).Again a(r) bounded between 0 and -1
is consistent
  with (31),(32) and (34).Thus for +ve topological charge the ansatz (29)
 with the following
  boundary conditions
  \begin {eqnarray}
  F(0) = 0\hspace{.2cm}	 a(0) = 0\nonumber\\
  F(\infty)={\pi \over 2} \hspace{.2cm}a(\infty)= -1
  \end {eqnarray}
  is consistent with the Bogomol'nyi conditions.
Similarly the consistency may be verified for -ve T.

  To conclude,we find that gauging the U(1) subgroup of the nonlinear O(3)
sigma model along with the inclusion of a self-interaction potential with
degenerate minima where the U(1) symmetry is spontaneously broken provides
topologically stable soliton solutions which have the desirable feature of
the removal of scale invariance with quantised energy,charge,flux and
angular momentum pertaining to each topological sector in contrast to [15]
where such solutions are infinitely degenerate.This breaking of the degeneracy
is ascribed to the topology of the minima of the potential considered in our
model.We have demonstrated that the theory satisfies Bogomol'nyi conditions
and discussed the consistency of the solutions.Detailed calculations of the
profiles are pending with other related issues.We propose to take up these
works subsequently.

 I like to thank Dr.R.Banerjee for helpful discussions and Dr.S.Roy
for his encouragements.I also thank Professor C.K.Majumdar,Director S.N.
Bose National centre for Basic Sciences for allowing me to use some
of his institute facilities.Finally my earnest thanks are due to the
referee for his comments which largely enabled me to put the work in
proper perspectives.


\begin{thebibliography}{99}
\bibitem [18] {kn:xx} Note that a factor 1/2 is missing in eqn(7) of [15].
\end{thebibliography}
 \end {document}